\documentclass{aastex63}

\usepackage{comment}

\received{}
\revised{}
\accepted{}
\submitjournal{ApJ}

\shorttitle{Accretion to non-spherical compact object}
\shortauthors{Datta et al.}

\graphicspath{{./}{figures/}}
\begin{document}

\title{Advective accretion onto a non-spherical accretor in white dwarf and neutron star binaries: a new scenario of shock formation}
\correspondingauthor{Sudeb Ranjan Datta}
\email{sudebd@iisc.ac.in}

\author{Sudeb Ranjan Datta}
\affiliation{Dept. of Physics, Indian Institute of Science, Bangalore-560012, India}

\author{Prasun Dhang}
\affiliation{Institute for Advanced Study, Tsinghua University, Beijing-100084, China}
\affiliation{Department of Astronomy, Tsinghua University, Beijing-100084, China}

\author{Bhupendra Mishra}
\affiliation{Los Alamos National Lab, Los Alamos, New Mexico 87545, United States}

\begin{abstract}

Numerous studies on hydrodynamics of the Keplerian as well as the sub-Keplerian accretion disc around a compact object (e.g., white dwarf (WD), neutron star (NS), or a black hole (BH)) attempted to explain the observed UV, soft and hard X-ray spectra. Although, when the compact object (e.g., a WD or an NS) has a finite surface, its rapid rotation, the stellar magnetic field could cause deformation of the spherical symmetry. Earlier studies for Keplerian disc showed that a deviation from the spherical symmetry of the compact object could affect the observed light curve and spectra at high frequencies. Here, we have explored the effect of the non-spherical nature of a compact object on the hydrodynamics of an optically thin, geometrically thick sub-Keplerian advective flow. We find that due to non-spherical shape of the central accretor, there is a possibility to trigger Rankine-Hugoniot shock in the sub-Keplerian advective flow close to the accretor without considering any general relativistic effect or presence of the hard surface of the star. Our results are more relevant for accretion onto WD as hardly any general relativistic effect will come in the picture. We propose that some observational features e.g., high significance of fitting the spectra with multi-temperature plasma models rather than single temperature models, and variable efficiency of X-ray emission (X-ray luminosity in comparison with optical and UV luminosity of the disk) in nonmagnetic cataclysmic variables can be explained by the presence of a shock in the sub-Keplerian advective flow.

\end{abstract}

\keywords{hydrodynamics; accretion, accretion disks; (stars:) novae, cataclysmic variables;  X-rays: binaries}

\section{Introduction} \label{sec:intro}

The physics of accretion discs has remained an open area of research since last few decades. Despite great progress in our understanding, we still need better theoretical model to shed light on recent observations. Generally, there are three theoretical models of accretion discs: (i) geometrically thin and optically thick Keplerian disc (\citealt{Shakura1973}), (ii) geometrically thick and optically thin advective flows (\citealt{Chakrabarti1989, Narayan1994}) and (iii) geometrically and optically thick slim discs (\citealt{Abramowicz1988}). Various accreting sources often show variability in observed luminosity and could change the geometrical shape of accretion discs.

Presence of a geometrically thick and optically thin advective flow is essential to explain hard X-rays and soft $\gamma$-rays from accreting systems like neutron stars (NSs) and black holes (BHs) (\citealt{Chakrabarti1995, Narayan1998, Yuan2005}). Due to the advective properties of thick flows, thermal energy generated due to viscous dissipation gets advected into central object before it can escape vertically from the disc surface. The advection of thermal energy in such geometrically thick hot flows plays an important role in stabilization against thermal instability. Flow remains hot, optically thin, geometrically thick and is able to give higher energy photons (\citealt{Narayan1994, Narayan1995}). On the other hand in a Keplerian, optically thick, geometrically thin Shakura-Sunyaev disc (SSD) (\citealt{Shakura1973, Novikov1973, Pringle1981}), emission of multi-temperature blackbody radiation is sufficient to balance the viscous heating. The resultant emission is observed in soft X-rays and is of higher luminosity. To explain the observed spectra, two-component accretion flow is also invoked (\citealt{Chakrabarti1995, Chakrabarti1996, Chatterjee2018}).

Several observations show that like BH and NS systems, accreting non-magnetic white dwarf systems (cataclysmic variables; CVs) also produce hard X-rays in all types of source states (and not soft X-rays as would be expected in high accretion rate states), the origin of which is interpreted to be connected to the presence of a radiatively inefficient (advective) accretion flow (\citealt{Balman2014, Godon2017, Balman2020} and references therein). However, \cite{Pringle1979} proposed the possibility of emission of hard X-rays from non-magnetic CVs under low accretion rate depending on effectiveness of shock in a Keplerian SSD. For a detail discussion on the presence of advective flows in non-magnetic CVs see section \ref{sect:adaf_cv}.

One crucial difference between the advective flow around the NS or the BH and that onto a WD is that protons are unable to transfer their energy to electrons and two temperature plasma comes into the picture for the former case (\citealt{Narayan1995, Rajesh2010}); while for the latter, temperature of the protons and the electrons remains almost the same, however see section (\ref{section for connecting shock with observables}) (\citealt{Medvedev2002, Frank2002}). Therefore, radiatively inefficient advective flow in CVs can be treated as a single temperature fluid (\citealt{Chakrabarti1989, Narayan1993, Popham1995, Chakrabarti1996}). For this preliminary model, we assume here that the disk is completely ionized.

It is probable that presence of the stellar rotation, strong stellar magnetic field and continuous accretion can change the spherical shape of the central accretor (\citealt{1968ApJ...151.1089O, 1968ApJ...153..797O, Shapiro1983, Komatsu1989, 2008MNRAS.385..531H, Das2015, Subramanian2015}). One first-order generalization of spherical shape is Maclaurin Spheroid (MS; \citealt{Chandrasekhar1969}). Deviation from the spherical symmetry of accreting source leads to change in gravitational force exerted on a test particle orbiting very close to it and could alter the dynamics of the accretion flow. Despite such an effect of gravity, the gravitational potential of an MS remains Newtonian and we do not consider any general relativistic (GR) effects for this work. 

There are already few theoretical studies on orbits around an MS (\citealt{Amsterdamski2002, Kluzniak2013}) and SSDs around an MS accretor (\citealt{Mishra2015}). These studies showed the emergence of many new features in the accretion flow just because of the deformation of the shape of the compact object. In this paper, we study the sub-Keplerian advective flow around an NS and a WD considering the accretor to be an  MS. To do a preliminary analysis, we consider a low angular momentum (sub-Keplerian) inviscid flow. It is convenient to assume such low value of angular momentum in modelling sub-Keplerian advective flow and its characteristic properties (\citealt{Chakrabarti1989, Das2001, Palit2019}). However, it must be emphasized that our reported model in this article is primarily applicable to WD since GR effects do not play a crucial role in accretion discs around such a compact object. The importance of our reported theory could also be realized to describe accretion flow around rapidly rotating NS but not very precisely without including all the GR effects. Therefore, in the rest of the article, we primarily focus on accretion discs around WD rather than NS. Although inclusion of GR effects will change the properties of the flow and will change the parameter space for our model to be applicable, we expect the qualitative picture will remain same.

Inner sonic point (Cf. Section \ref{formalism}) does not exist for accretion onto a spherical Newtonian accretor. That is why shock can not be realized within the sub-Keplerian advective flow around a spherical Newtonian accretor unless presence of hard surface (\citealt{Dhang2016}) or GR correction in gravitational potential is taken into account (\citealt{Chakrabarti1989, Chakrabarti1996, Mukhopadhyay2002, Dihingia2020}). For accretion  on to a WD, to the best of our knowledge, we are reporting for the first time that formation of a shock is possible in the appropriate parameter space due to the deformed shape of the accretor. To some extent, the shocks can explain the scenarios where the hard X-ray emission components that are predominantly detected both in the quiescence and outburst in the inner disk and the X-ray emitting region. There seem to be a few detections of the Soft X-ray regime during outbursts (i.e., state transition to high state) and no detection of Soft X-ray emission in persistent high state CVs. This suggests that an optically thick standard boundary does not form (see discussions in \citealt{Balman2014}, \citealt{Balman2020}[sec 2.3] and references therein). The shocks we calculate, can explain the detected hard X-ray emissions in non-magnetic CVs when the different parameters of the central accretor and of the flow are suitable for the shocks to occur.

The plan of the paper is as follows. In section 2, we present the formalism we follow. We analyze the parameter space for advective accretion flow in section 3. In section 4, we present the hydrodynamics of flow around an NS as well as a WD. We discuss the possible connection of our work to observations in section 5. Finally we concluded in section 6. 

\section{A general view of the formalism}
\label{formalism}
We use a standard approach (e.g. see \citealt{Chakrabarti1989, Mukhopadhyay2003}) to investigate the inviscid accretion flow around a compact star (e.g. a neutron star or a white dwarf) whose gravity can be described by the gravitational force due to an MS. Specifically, we want to study the effects of deformation (may be due to rotation of the star or due to the intrinsic stellar magnetic field).
   
To obtain the flow variables, we solve conservation of mass (mass continuity equation) and momentum (radial momentum balance equation) as given below-
\begin{equation}\label{continuity equation}
\frac{d}{dr}(r\Sigma v)=0,
\end{equation} 
\begin{equation}\label{radial momentum balance}
v\frac{dv}{dr}+\frac{1}{\rho}\frac{dP}{dr}-\frac{l^2}{r^3}+F(r)=0,
\end{equation} 
where $\Sigma=2\rho(r)h(r)$ is the vertically integrated surface density and $F(r)$ is the gravitational force for MS. Here, $\rho(r)$ is the density and $h(r)$ is the half-thickness of the disc. Unless otherwise stated, all the radial distances in this paper are in units of $GM$/$c^2$, where $M$ is the mass of the MS, $G$ is the gravitational constant and $c$ is the speed of light. The velocity $v$ is in units of $c$ and specific angular momentum $l$ is in units of $GM/c$.

Along with equations (\ref{continuity equation}) and (\ref{radial momentum balance}), we use polytropic equation 
\begin{equation}\label{equation of state}
P=K\rho^{\gamma}
\end{equation}
to describe the equation of state. Where $\gamma$ is the adiabatic index. Gravitational force due to MS at equatorial plane is
\begin{equation}\label{force MS}
F(r)=\Omega^2 r=2\pi G \rho_{\rm *} (1-e^2)^{1/2}e^{-3}(\gamma_r-\cos\gamma_{\rm r}\sin\gamma_{\rm r})r
\end{equation} 
where $e$ is the eccentricity and $\rho_\mathrm{*}$ is the density of MS, taken to be uniform. $\gamma_\mathrm{r}$=sin$^{-1}(ae/r)$, $a$ is the semi-major axis of MS.

In dimensionless units, force takes the form
\begin{equation}\label{force MS dimensionless}
F(r)=1.5(ae)^{-3}(\gamma_{\rm r}-\cos\gamma_{\rm r}\sin\gamma_{\rm r})r
\end{equation} 

and $h(r)$ can be found from the vertical equilibrium equation as 
\begin{equation}
h(r)=c_{\rm s}r^{1/2}F^{-1/2}
\end{equation} 
where $c_\mathrm{s}=\sqrt{\gamma p/\rho}$ is the sound speed. Unless mentioned specifically, we use $\gamma$=4/3.

 Now, combining equation (\ref{continuity equation}) and (\ref{radial momentum balance}) we get 
\begin{equation}\label{slope of velocity}
\frac{dv}{dr}=\left[\frac{l^2}{r^3}-F(r)+\frac{c^2_{\rm s}}{\gamma+1}\left(\frac{3}{r}-\frac{1}{F}\frac{dF}{dr}\right)\right]\Bigg/\left[v-\frac{2c^2_{\rm s}}{(\gamma+1)v}\right]
\end{equation} 

Far away from the accretor, the radial velocity $v$ is small and subsonic. As the flow approaches the accretor, $v$ increases and in principle can exceed the local sound speed $c_\mathrm{s}$ i.e. can become supersonic. This implies that at a critical radius $r_\mathrm{c}$, the denominator of equation (\ref{slope of velocity}) becomes zero i.e. at $r_\mathrm{c}$, $v$ becomes equal to $\sqrt{2/(\gamma+1)} c_\mathrm{s}$. Though $v \neq c_\mathrm{s}$ at $r=r_\mathrm{c}$, historically, a standard trend in literature is to write subsonic and supersonic flow before and after the critical point ($r_\mathrm{c}$). We follow the same traditional nomenclature in this work also. For a realistic flow, the presence of a non-divergent velocity gradient leads to the requirement that the numerator of equation (\ref{slope of velocity}) also vanishes so that we can use L'Hospital rule to have a definite $\frac{dv}{dr}\Bigg|_c$.  Using L'Hospital's rule, the slope for velocity at critical point becomes
\begin{equation}\label{slope of velocity at critical point}
\frac{dv}{dr}\Bigg|_{\rm c}=-\frac{B+\sqrt{B^2-4AC}}{2A}
\end{equation}

where 
\begin{equation}
A=1+\frac{2c^2_{\rm sc}}{(\gamma+1)v^2_{\rm c}}\left[1+2\left(\frac{\gamma-1}{\gamma+1}\right)\right],
\end{equation}

\begin{equation}
B=\frac{4c^2_{\rm sc}(\gamma-1)}{v_{\rm c}(\gamma+1)^2}\left(\frac{3}{r_{\rm c}}-\frac{1}{F_{\rm c}}\frac{dF}{dr}\Bigg|_{\rm c}\right),
\end{equation}

\begin{equation}
C=\frac{3l^2}{r^4_{\rm c}}+\frac{dF}{dr}\Bigg|_{\rm c}+\frac{c^2_{\rm sc}(\gamma-1)}{(\gamma+1)^2}\left[\frac{3}{r_{\rm c}}+\frac{1}{F_{\rm c}}\frac{dF}{dr}\Bigg|_{\rm c}\right]^2-\frac{c^2_{\rm sc}}{\gamma+1}\left[\left(\frac{1}{F_{\rm c}}\frac{dF}{dr}\Bigg|_{\rm c}\right)^2
 -\frac{1}{F_{\rm c}}\frac{d^2F}{dr^2}\Bigg|_{\rm c}-\frac{3}{r^2_{\rm c}}\right]
\end{equation}

Equating denominator and numerator of (\ref{slope of velocity}) to zero, we get Mach number
\begin{equation}
\mathcal{M}_{\rm c}=\frac{v_{\rm c}}{c_{\rm sc}}=\sqrt{\frac{2}{\gamma+1}}
\end{equation}
and sound speed
\begin{equation}
c_{\rm sc}=\sqrt{(\gamma+1)\left(F_{\rm c}-\frac{l^2}{r^3_{\rm c}}\right)\Bigg/\left(\frac{3}{r_{\rm c}}-\frac{1}{F_{\rm c}}\frac{dF}{dr}\Bigg|_{\rm c}\right)}
\end{equation}
at the critical point $r_\mathrm{c}$.
Now integrating equations \ref{continuity equation} and \ref{radial momentum balance}, we can write down the energy and entropy of the flow at critical point as 
\begin{equation}
E_c=\frac{2\gamma}{(\gamma-1)}\left[\left(F_{\rm c}-\frac{l^2}{r^3_{\rm c}}\right)\Bigg/\left(\frac{3}{r_{\rm c}}-\frac{1}{F_{\rm c}}\frac{dF}{dr}\Bigg|_{\rm c}\right)\right]+V_{\rm c}+\frac{l^2}{2r^2_{\rm c}}
\end{equation}
\label{energy at critical point}
and
\begin{equation}
\dot{\mu}_{\rm c}= (\gamma K)^n\dot{M}=r^{3/2}_{\rm c}F^{-1/2}_{\rm c}(\gamma+1)^{q/2} \times \left[\left(F_{\rm c}-\frac{l^2}{r^3_{\rm c}}\right)\Bigg/\left(\frac{3}{r_{\rm c}}-\frac{1}{F_{\rm c}}\frac{dF}{dr}\Bigg|_{\rm c}\right)\right]^{\gamma/(\gamma-1)}
\end{equation}

where $V_\mathrm{c}$=($\int_{}{}Fdr)|_\mathrm{c}$, $q=(\gamma+1)/[2(\gamma-1)]$ and $n=1/(\gamma-1$) . Here $\dot{\mu}$ carries the information of entropy. For non-dissispative system $E$ remains constant throughout the flow. Here an important point to note is that formation of a shock enhances the entropy which changes the value of `$K$' which finally changes the value of $\dot{\mu}_\mathrm{c}$. Therefore, due to mass continuity $\dot{M}$ remains same throughout the flow, while formation of a shock increases the value of $\dot{\mu}_\mathrm{c}$ due to increment of entropy.

\begin{figure}

\includegraphics[scale=0.75]{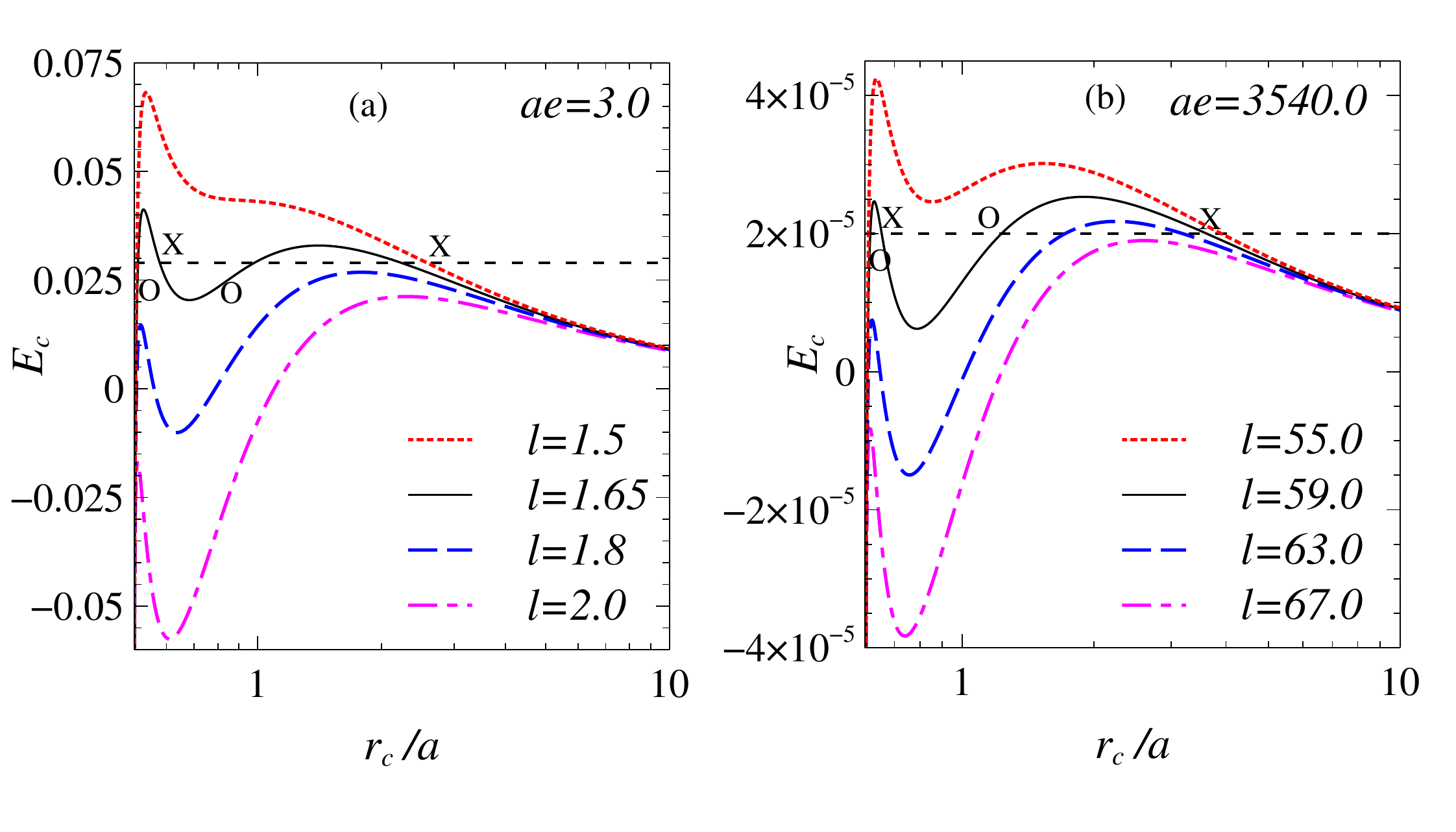}
\caption{Variation of energy as a function of critical point location for various values of specific angular momentum ($l$). Left panel (a) is for parameters for a typical NS. $ae$ is fixed at 3.0 where $a$ and $e$ are assumed to be 6.0 and 0.5 respectively. The horizontal line indicates constant energy of 0.029. Right panel (b) is for a typical WD. $ae$ is fixed at 3540.0 where $a$ and $e$ are assumed to be 5900.0 and 0.6 respectively. The horizontal line indicates constant energy of $2.0\times10^{-5}$. Adiabatic index $\gamma$=4/3. Different types of critical points are marked in the figure. The x-axis values are presenting critical point location in terms of semi-major axis ($a$) of the compact object.}
\label{lambda sets}
\end{figure}

\begin{figure}

\includegraphics[scale=0.75]{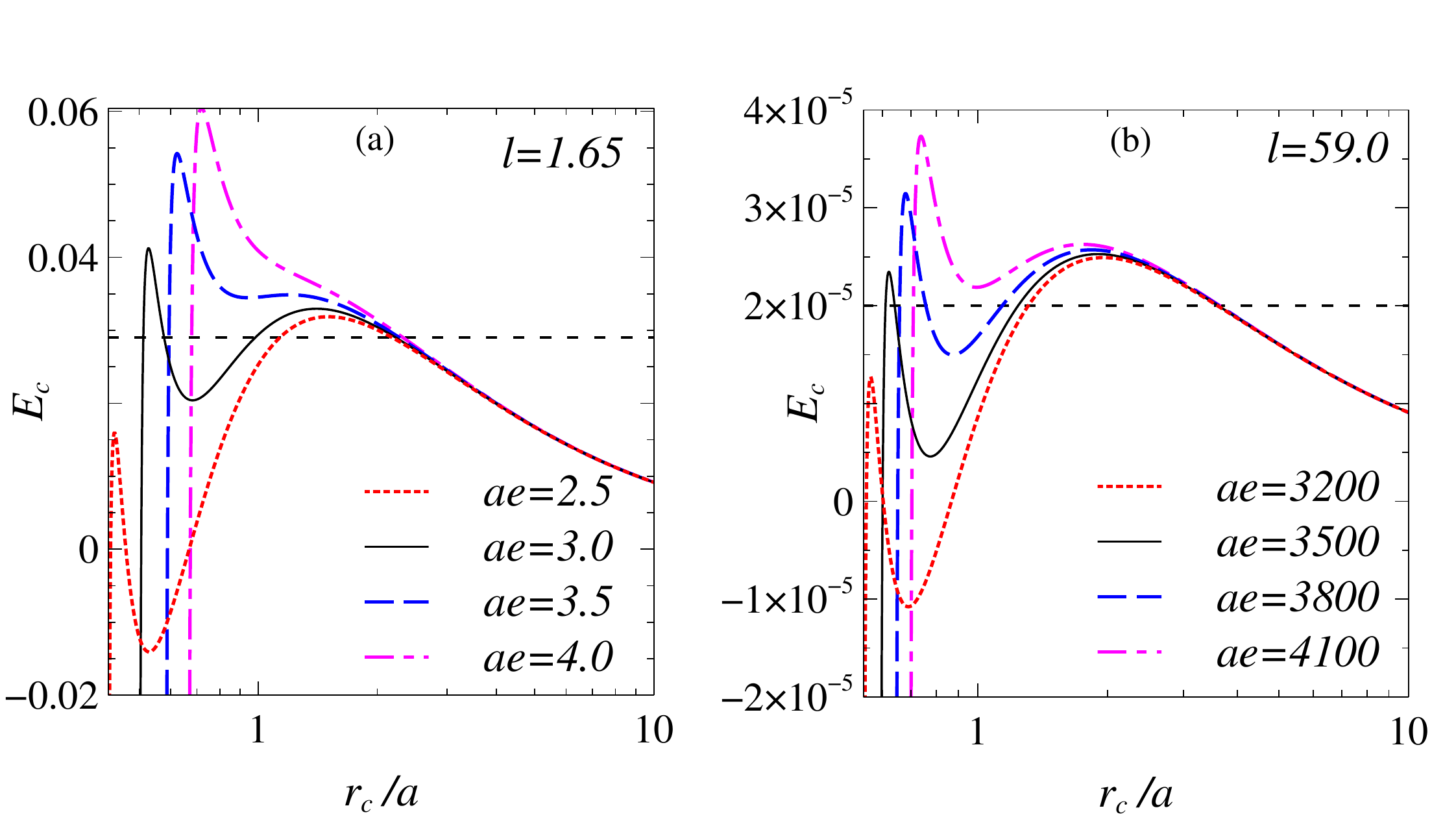}
\caption{Variation of energy as a function of critical point location for different values of (semi-major axis($a$) $\times$ eccentricity($e$)). Left panel (a) is for parameters for a typical NS. $l$ is fixed at 1.65. The horizontal line indicates constant energy of 0.029. Right panel (b) is for a typical WD. $l$ is fixed at 59.0. The horizontal line indicates constant energy of $2.0\times10^{-5}$. Adiabatic index $\gamma$=4/3. The x-axis values are presenting critical point location in terms of semi-major axis ($a$) of the compact object where $a$ is assumed to be 6.0 for NS and 5900.0 for WD. Keep in mind that only whole value of $a\times e$ affects the dynamics as well as energy, not their individual values.}

\label{ae sets}
\end{figure}

Fig. $\ref{lambda sets}$ and $\ref{ae sets}$ show the variation of energy at critical point ($E_\mathrm{c}$) with position of critical point ($r_\mathrm{c}$) for different $l$ and $ae$ values respectively for the parameters applicable to a typical NS and WD. As we investigated here non-dissipative flow, so, fixing energy at critical point will fix the energy of the flow. Horizontal line in the Fig. $\ref{lambda sets}$ and $\ref{ae sets}$ indicates the constant energy of the flow. It is clear that depending on $E_\mathrm{c}$, critical points' locations as well as number of critical points also varies.
 
This opens up another possibility of formation of Rankine-Hugoniot shock (\citealt{landau1987fluid}) in the flow. If we generalize the conditions to form a shock in accretion disc given by \cite{Chakrabarti1989} for any compact object, we get
 
\begin{equation}\label{shock_1}
\frac{1}{2}\mathcal{M}^2_{\rm +}c^2_{\rm s,sh+}+nc^2_{\rm s,sh+}=\frac{1}{2}\mathcal{M}^2_{\rm -}c^2_{\rm s,sh-}+nc^2_{\rm s,sh-}
\end{equation} 

\begin{equation}\label{shock_2}
\frac{c^{\nu}_{\rm sh+}}{\dot{\mu}_{\rm +}}\left(\frac{2\gamma}{3\gamma-1}+\gamma \mathcal{M}^2_{\rm +}\right)=\frac{c^{\nu}_{\rm sh-}}{\dot{\mu}_{\rm -}}\left(\frac{2\gamma}{3\gamma-1}+\gamma \mathcal{M}^2_{\rm -}\right)
\end{equation}

\begin{equation}\label{shock_3}
\dot{\mu}_{\rm +}>\dot{\mu}_{\rm -}
\end{equation}

where 
\begin{equation}\label{shock_4}
\dot{\mu}=\mathcal{M}c^{2(n+1)}_{\rm s}\frac{r^{3/2}_{\rm s}}{\sqrt{F(r_{\rm s})}}
\end{equation}

Here, the subscript `$sh$' indicates the shock, `-' and `+' subscripts indicate before and after the shock. $\mathcal{M}$ denotes the Mach number of the fluid and exponent of $c_s$ in equation (\ref{shock_2}) is $\nu$=(3$\gamma$-1)/($\gamma$-1). Equation (\ref{shock_3}) indicates the natural choice of formation of shock as entropy increases after the shock. Combining equations (\ref{shock_2}) and (\ref{shock_4}), we get the shock invariant quantity as 

\begin{equation}\label{shock invariant quantity}
SIQ=\frac{[2/\mathcal{M}_{\rm +}+(3\gamma-1)\mathcal{M}_{\rm +}]^2}{\mathcal{M}^2_{\rm +}(\gamma-1)+2}=\frac{[2/\mathcal{M}_{\rm -}+(3\gamma-1)\mathcal{M}_{\rm -}]^2}{\mathcal{M}^2_{\rm -}(\gamma-1)+2}
\end{equation}

If conditions (\ref{shock_1}), (\ref{shock_3}) and (\ref{shock invariant quantity}) are satisfied simultaneously by the flow, then shock will form in accretion disc. Though actual values will vary from source to source, due to this shock within advective flow, typically the radial velocity is almost halved. From mass continuity, due to decrement in radial velocity, the density doubles in the post-shock region which will enhance the cooling in comparison with pre-shock flow.

\section{Analysis of parameter space}
\label{sect:param_space}
In this section, we discuss the dependence of flow properties on different parameters such as eccentricity ($e$), radius of the central accretor ($a$), and specific angular momentum of the flow ($l$). Fig. $\ref{lambda sets}$ shows how the variation of specific angular momentum of the flow changes the number as well as location of critical points (\citealt{Chakrabarti1989}) for fixed values of energy $E_\mathrm{c}$ and $ae$. For example, if we fix $ae$=3.0 (3540.0 for WD case) as well as the energy of the flow at 0.029 ($2.0\times10^{-5}$) then for $l=1.65$ ($l=59.0$), there are four critical points (where horizontal line of value 0.029 ($2.0\times10^{-5}$) cuts $E_\mathrm{c}$ vs. $r_\mathrm{c}$ curves). The outer most critical point is a `X' type critical point (Slope of $E_\mathrm{c}$ vs $r_\mathrm{c}$ curve is negative). As we move inside, next critical points are `O' type, `X' type and `O' type respectively. \cite{Chakrabarti1990} discusses critical points and their types in detail. For the present purpose it is sufficient to state that `X' type critical points are responsible for successful accretion. So, our current interest lies in the outer and inner `X' type critical points. Hereafter we indicate these two `X' type critical points as outer and inner critical points. Changing the specific angular momentum of the flow will change $E_\mathrm{c}$ of the flow according to equation (\ref{energy at critical point}) as shown by the different curves in Fig. $\ref{lambda sets}$. Now, for a specific case, the energy of the flow will be fixed for an inviscid flow, which is indicated as a straight line in Fig. $\ref{lambda sets}$. Therefore, for a particular value of $E_\mathrm{c}$ (0.029 for a typical NS and $2.0\times10^{-5}$ for a typical WD) and $ae$ (3.0 for a typical NS and 3540.0 for a typical WD), only a definite range of specific angular momentum ($l$ $\sim$ 1.6-1.7 for accretion onto a typical NS and $l$ $\sim$ 58.0-59.0 for accretion onto a typical WD) can give rise simultaneous occurrence of outer and inner critical points, which is shown in Fig. $\ref{lambda sets}$. For other $l$ values, with the same $E_\mathrm{c}$ and $ae$, there are different possibilities of occurrence of one `X' type and one `O' type or one `X' type or no `X' type critical point. The consequence of different possibilities are discussed in detail in \cite{Chakrabarti1990}. However, our current interest lies in the possibility of shock in the flow which is possible only when two `X' type critical points are present simultaneously. 

If the angular momentum is low enough, the centrifugal barrier is unable to create any characteristic feature in the flow i.e. occurrence of multiple `X' type critical points is not possible and flow is smoothly accreted by the accretor. This indicates that due to gradual decrement of specific angular momentum ($l$), the flow becomes Bondi-like (\citealt{Bondi1952}) as angular momentum is unable to affect the flow. It is clear from Fig. $\ref{lambda sets}$ for $l \leq ~1.5$ in case of an NS and $l \leq ~52.0$ in case of a WD, the flow becomes Bondi-like for our chosen parameter space as $E_c$ increases monotonically with decreasing $r_c$ due to disappearance of effective centrifugal barrier. It should be mentioned that the presence of both outer and inner critical points is necessary to make the shock in the flow probable (\citealt{Chakrabarti1989}).

Next, we show how the change in eccentricity and radius affects critical point analysis. Treating central accretor as MS, fundamental difference is coming through the gravitational force it is applying on the test particle (Equation (\ref{force MS dimensionless})) instead of spherical accretor. As this force does not involve $a$ and $e$ separately, changing the value of product of $a$ and $e$ only affects the results, not the individual values of $a$ and $e$. Fig. $\ref{ae sets}$ shows the $E_\mathrm{c}$ vs. $r_\mathrm{c}$ plots for different values of $ae$ and a fixed $l$. Fig. $\ref{ae sets}$ shows that for a fixed $E_\mathrm{c}$ (0.029 for a typical NS and $2.0\times10^{-5}$ for a typical WD) and $l$ (1.65 for accretion onto a typical NS and 59.0 for accretion onto a typical WD), small range of $ae$ ($\sim$ 2.976-3.018 for NS and $\sim$ 3451.5-3599.0 for WD) can make the shock possible, as like Fig. $\ref{lambda sets}$, which shows the range of $l$ to make the shock happen. These two figures together indicate that when a flow with fixed energy accretes to a suitable accretor (appropriate value of $ae$) with suitable specific angular momentum value, shock will occur. One set of values of different parameters suitable for shock for an NS having $a$=6.0 and for a WD having $a$=5900.0 is tabulated in Table (\ref{range for shock}).

Equation $(\ref{force MS dimensionless})$ indicates that for a realistic $F(r)$, $r \ge ae$. Therefore, while doing a parameter space survey, we restrict to the limit $r \ge ae$ value. However, when we present the hydrodynamics of the accretion flow in the next section, we assume that disc truncates at the surface of the star of radius $a$. It may arise that the inner critical point lies inside $a$. In that case, we do the hydrodynamics considering the inner critical point and truncate the accretion flow at $r=a$ to mimic the presence of the surface.

\begin{figure*}

\includegraphics[scale=0.7]{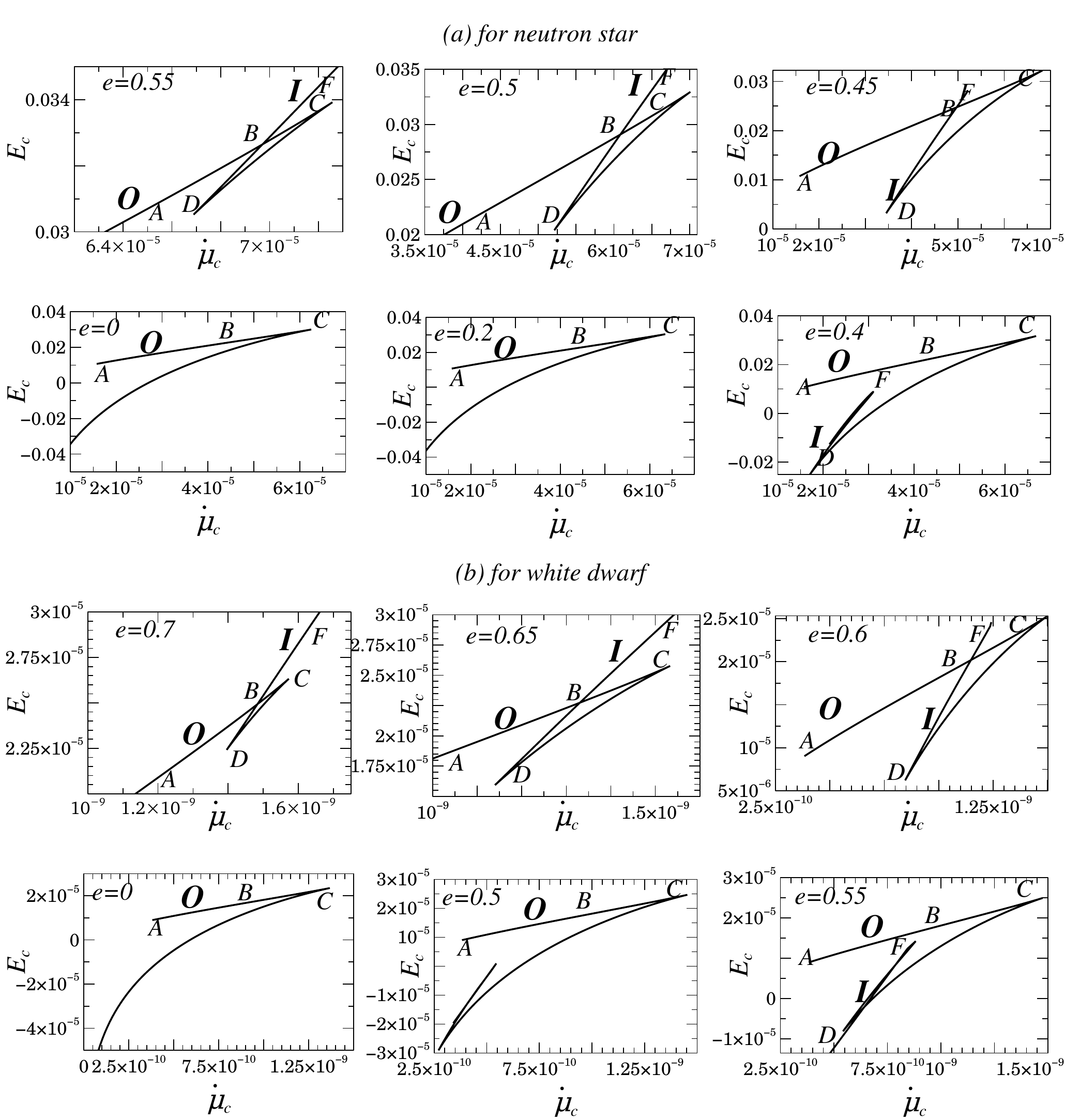}
\caption{Variation of possibility of shock with different eccentricity. (a) for an NS case $a$=6.0 and $l$=1.65 and (b) for a WD case $a$=5900.0 and $l$=59.0 are fixed. $\bold{O}$ (AC line) and $\bold{I}$ (FD line) are representing outer and inner critical branches respectively. The range of energy value for which $E_c$ = constant line passes through both the branches simultaneously gives the possibility of shock formation. It shows that decreasing $e$ increases the possibility of shock up to a certain value for which shock can occur i.e. till when two branches exist simultaneously.}

\label{Ec meuc with eccentricity}
\end{figure*}

For a transonic flow around an accretor with a hard surface, shock in the accretion flow is almost inevitable because flow has to slow down at the surface (\citealt{Dhang2016, Dhang2018}). Possibility of formation of the shock enhances in an inviscid sub-Keplerian transonic flow if both the outer and inner critical points coexist (\citealt{Chakrabarti1989, Chakrabarti1996, Mukhopadhyay2003}). In this work, we do not consider the effects of presence of the surface and assume there is a sink of mass at the inner boundary. To investigate the formation of shock in principle, we can solve the hydrodynamics from inner critical point and from outer critical point and can check whether the shock conditions (equations (\ref{shock_1})-(\ref{shock invariant quantity})) are satisfied or not in the region between two critical points. However, we follow in a more convenient way to check the possibility of shock which is by plotting $E_\mathrm{c}$ vs. $\dot{\mu}_\mathrm{c}$ by varying $r_\mathrm{c}$ as shown in Fig. \ref{Ec meuc with eccentricity}. This is usually called `swallow-tail' picture (Fig. \ref{Ec meuc with eccentricity}). For accretion with shock in the current context, the picture is like the following: matter from companion which is coming from far away reaches the outer critical point, becomes supersonic and then due to shock it jumps to the inner critical point branch and finally accretes onto the compact object. In other scenarios, it may be possible to jump to other branches or shock may not happen and matter directly accretes to the compact object. For the present purpose we restrict ourselves to the scenario to jump to inner critical branch due to the formation of shock.

`Swallow-tail' picture gives the possible range of energy for which shock may form. In  Fig. \ref{Ec meuc with eccentricity}, `O' and `I' represent the branch for outer and inner critical point respectively. We need to find energy value for which matter can jump from the outer critical branch to inner critical branch with the necessary condition of increment of entropy as given in equation (\ref{shock_3}). This means, in Fig. \ref{Ec meuc with eccentricity}, matter will jump from AB line (branch corresponding to the  outer critical point with lower entropy) to DB line (branch corresponding to the inner critical point with a higher entropy) to make the shock happen. If we draw a constant energy line (which is the energy of the flow) which passes through AB and DB line simultaneously, then matter with that energy may jump from low entropy region to high entropy region keeping its energy constant. As the probability increases for a y=constant line to intersect AB and DB simultaneously, probability of occurrence of shock increases. Keep in mind that it only gives the probability of the shock. Getting a rough estimate of the energy range with other fixed parameters for the shock, we need to check whether other condition (equation (\ref{shock invariant quantity})) of the shock is satisfying or not.\\

Fig. \ref{Ec meuc with eccentricity} shows how the eccentricity of the MS changes the possibility of the occurrence of shocks for a typical (a) NS and (b) WD. We found that with decreasing eccentricity keeping the $l$ value fixed, the probability for a y=constant line to cut AB and DB line simultaneously increases, which indicates that decreasing eccentricity increases the possibility of formation of shock. However, for NS (WD) case $e$=0.4 (0.55) and $l$=1.65 (59.0), the curve from inner branch turns back and does not intersect with the outer branch. This returning nature is coming due to the presence of `O' type critical point in the innermost region as described earlier. In the rest of the swallow-tail diagrams also, the curve turns back from the inner branch after some value. To keep the picture easy to understand, we have plotted up to that point beyond which returning occurs. When $a$=6.0 (5900.0), $l$=1.65 (59.0), for $e$=0.4 (0.5) or lower values, there is no possible energy value which can intersect outer as well as inner branch. This shows that for the  range of “e” equal or smaller than a certain value (while other parameters remain fixed), inner and outer critical points can not exist simultaneously. According to the present scenario of shock formation, shock will not form. From this analysis we can say that decreasing eccentricity of MS increases the probability of shock as long as the outer and inner critical points exist simultaneously. For slowly rotating non-magnetic CVs eccentricity is small which indicates that occurrence of shock will be more probable unless inner critical point vanishes. Fig. \ref{Ec meuc with eccentricity} shows for a NS (WD) accretor, flow with specific angular momentum $l$ = 1.65 (59.0), accreted onto central accretor having $a$ = 6.0 (5900.0) with $e$ in the range 0.45-0.55 (0.55-0.7) is highly probable to make the shock happen.

\begin{figure}

\includegraphics[scale=0.8]{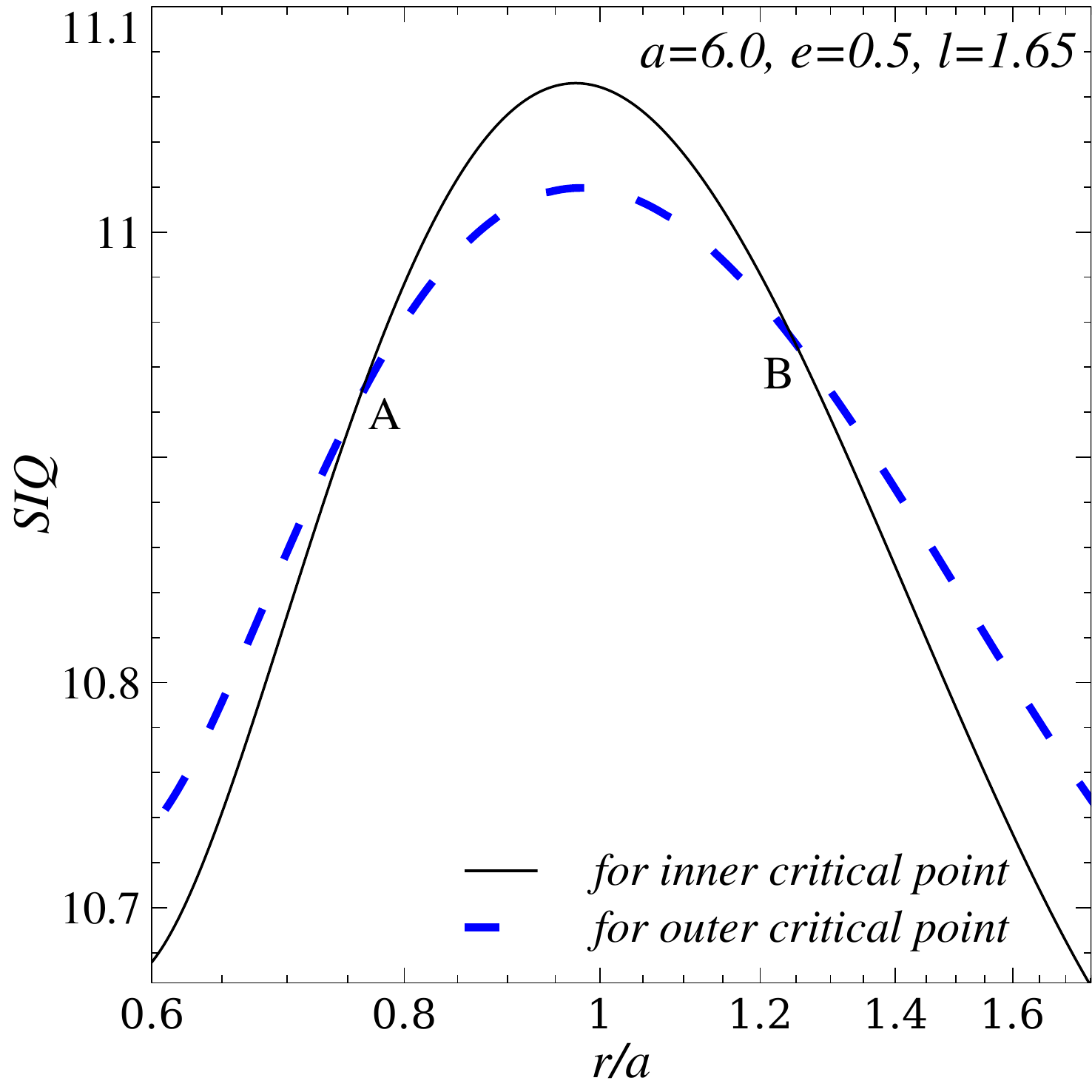}
\caption{Variation of shock invariant quantity with radial distance for accretion flow with $E_\mathrm{c}$=0.0289 and $l$=1.65 around neutron star. Two curves are for accretion through outer and inner critical point. The intersection points (A and B) of the two curves indicate the possible shock locations where inner one (A) is unstable and outer one (B) is representing the actual location of shock}

\label{SIQ}
\end{figure}

\section{Hydrodynamics around compact object}
\label{sect:hydrodynamics}
In this section, we investigate the accretion structure around the two different kinds of compact objects namely NSs and WDs whose shape can be described by the MS. However, it should be mentioned that while the used Newtonian approach can explain the accretion flow around a WD, general relativistic effects are significant to consider in case of accretion disc around NS. That is why our results presented here for an NS is not realistically accurate. However, we expect that the qualitative nature of the flow will remain same. 
 
\subsection{Around an NS}
\label{sect:ns_hydrodynamics}

Here we describe the hydrodynamics of the accretion flow assuming the neutron star as an MS with typical observed values of mass 1$M_\odot$, $a$=6.0. We have shown three cases depending on different realistic values of eccentricity ($e$) of NS and specific angular momentum ($l$) of the flow. In perfect combination of $e$ and $l$, shock forms, but in other cases it disappears.
 
In section 3, we see that shock can form for a specific range of parameter space ($l, e$). A convenient way to check whether shock forms or not is to check the equality in equation (\ref{shock invariant quantity}). To study it, we simultaneously plot $SIQ$ for two branches: when accretion happens through inner critical point and through outer critical point. Points of intersection give the locations of shock where flow jumps from outer critical branch to inner critical branch. This is shown in Fig. \ref{SIQ} for $l$=1.65 and $e$=0.5 when the $E_\mathrm{c}$ of the flow is 0.0289. Out of two possible shock locations, inner one is unstable under radial perturbation due to post-shock acceleration. The outer one is stable due to post-shock deceleration (\citealt{Nakayama1992}; \cite{Nobuta1994}). So, the outer intersection point (B) is our desirable shock location. However, even the outer shock location is unstable to non-radial, non-axisymmetric perturbation (\citealt{Iwakami2009}). These instabilities (oscillations) are invoked to explain different time variabilities (i.e. ejections, QPOs) in the accreting systems (\citealt{Molteni1996, Molteni1999, Bhattacharjee2019}).
 
We consider both shocked and shock-free solutions. This is achieved by solving the equations for $\frac{dv}{dr}$ and $\frac{dc_\mathrm{s}}{dr}$ simultaneously for different sets of $l$ and $e$ and keeping $E_\mathrm{c}$=0.0289 fixed. We specify flow parameters at the critical points (where quantities are well determined, see section 2), and look for the stationary solutions.
 
\begin{figure*}

\includegraphics{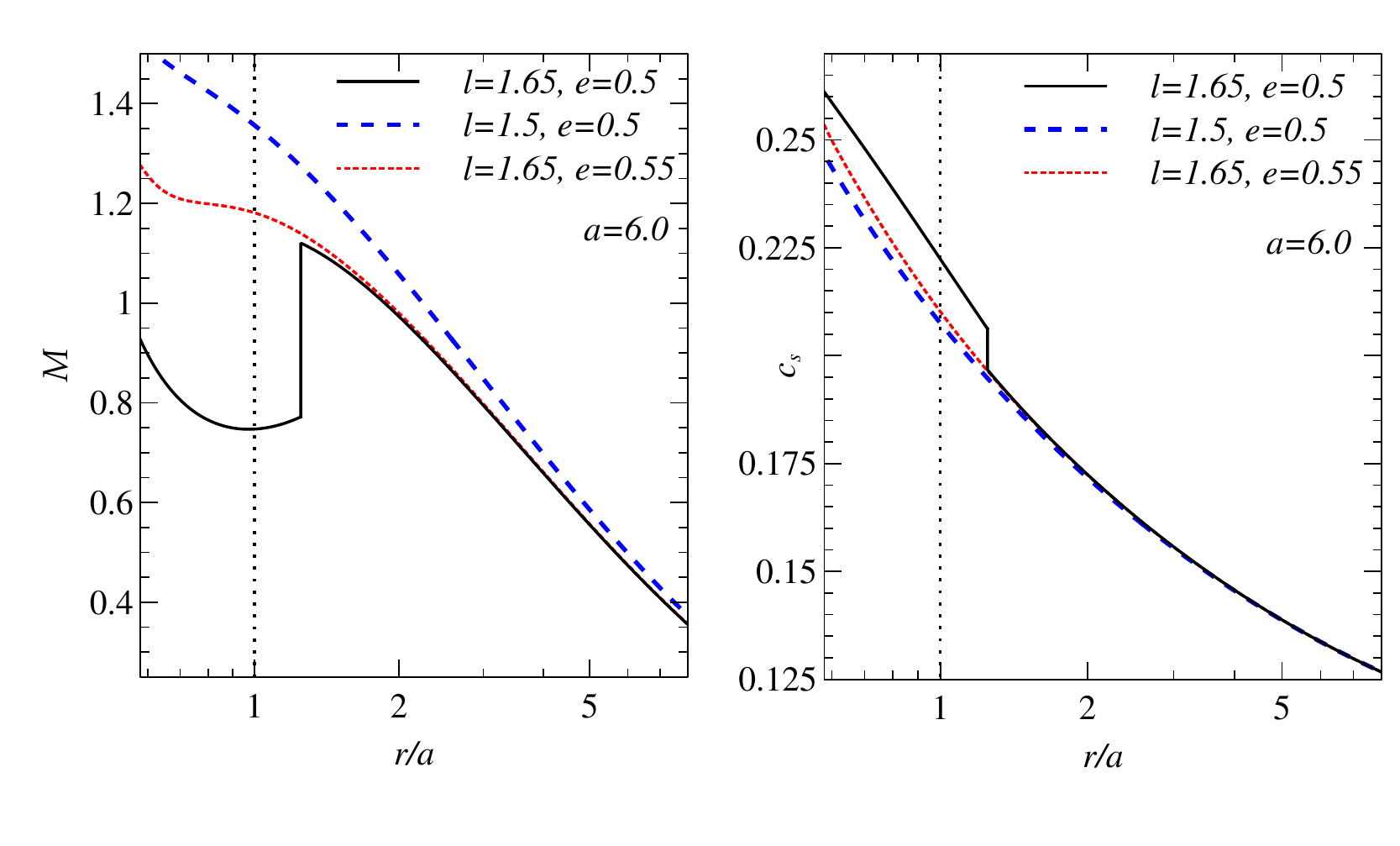}
\caption{Variation of Mach number($\mathcal{M}$) and sound speed($c_\mathrm{s}$) with radius with and without formation of shock for different combination of $l$ and $e$. The x-axis is presenting radius in terms of $a$. Combination of $l$=1.65 and $e$=0.5 gives $r_\mathrm{{c,in}}/a$=0.58 and $r_\mathrm{{c,out}}/a$=2.24 for $E_\mathrm{c}$=0.0289 which makes shock possible at $r_\mathrm{sh}/a$=1.25. The plot is extended till inner critical point for shock solution. The vertical dotted line at $r/a$=1 indicates the surface of the neutron star.}
\label{NS hydro with without shock}
\end{figure*}

Fig. \ref{NS hydro with without shock} shows the radial profiles of the Mach number ($\mathcal{M}$=$v/c_\mathrm{s}$) and sound speed ($c_\mathrm{s}$) for the accretion flow around NS for $E_\mathrm{c}$=0.0289, $a$=6.0 for three different sets of parameters: $l$=1.65, $e$=0.5; $l$=1.5, $e$=0.5; $l$=1.65, $e$=0.55. Although fixing $E_\mathrm{c}$ value to 0.0289 gives inner and outer critical point for set ($l$=1.65, $e$=0.5), same energy value gives only outer critical point for ($l$=1.5, $e$=0.5) and ($l$=1.65, $e$=0.55) which does not allow the formation of a shock. For the shock solution, the outer and inner critical point occurs at $r/a$=2.24 and 0.58 which makes the shock possible at $r/a$=1.25. Although, the inner critical point lies within the NS, the flow will remain same as the shock occurs within the flow outside of NS surface. Therefore, only ($l$=1.65, $e$=0.5) gives us shock in advective flow among the three sets. It indicates that accreting matter of the same energy may or may not form shock depending on its specific angular momentum and eccentricity of the central accretor. At the shock, there is a sudden jump in Mach number as well as in sound speed, whereas there is a smooth increase in $ \mathcal{M}$ and $c_\mathrm{s}$ for other cases as gradually accreted by neutron star. Fig. \ref{NS hydro with without shock} is extended till the inner critical point for shock solution. The vertical dotted line indicates the location of the surface where the flow will stop. One set of values of different parameters suitable for shock for an NS having $a$=6.0 is tabulated in Table (\ref{range for shock}). Once values of $a$, $e$, $l_c$ are fixed, there is range in $E_c$ for which shock will be possible. Similarly, when $E_c$ is fixed, there is a range in values of other parameters for which shock will be possible. However, it is evident as well as worth to mention that if we can vary all the parameters instaed of fixing two as tabulated in Table (\ref{range for shock}), the range of different parameters becomes quite large for which shock is possible. For accretion onto an NS with $a$=6.0, shock will be possible by tuning the values of different parameters from range of $e$=0.2-1.0, $l_c$=1.1-2.45 and $E_c$=0.005-0.090. Values of any parameter lying outside of this range will not be able to make the shock happen.   

In literature, the occurrence of shock is possible due to either presence of hard surface of the accretor (\citealt{Dhang2016}) or the GR correction in the gravitational potential (\citealt{Chakrabarti1989, Chakrabarti1996, Mukhopadhyay2002, Dihingia2020}). The beauty of the MS potential is in formation of the shock without incorporating any kind of general relativistic effect. Simultaneous formation of outer as well as inner critical points allows the formation of shock which is not possible using Newtonian potential of a spherical body. To include the GR effect approximately, it is customary to replace Newtonian potential by pseudo-Newtonian potential \cite{Paczynski1980} to mimic GR effects on the accretion flow. It should also be mentioned that the formation of shock in MS potential is possible even at lower $l$ values compared to what is expected when pseudo-Newtonian potential is used for the same $E_\mathrm{c}$. In addition to the MS potential effect, inclusion of GR effects for NS accretion will change the properties of the flow and will change the parameter space for our model to be applicable, we expect the qualitative picture i.e. the change in velocity as well as in sound speed due to shock, the increment in density as well as the enhancement in cooling and change in temperature will remain almost same.

\begin{table}
	\centering
	\caption{One set of values of different parameters for which shock will occur for accretion onto a typical NS and WD. Once values of $a$, $e$, $l_c$ are fixed, there is a range of $E_c$ for which shock will occur. This is true for other parameters also. Always there is a range in value of one parameter when other parameters are fixed.}
	\label{range for shock}
	\begin{tabular}{|l|c|c|c|r|}
	\hline
	central accretor & $a$ ($R_g$) & $e$ & $l_c$ & $E_c$\\
	\hline
	 &  & 0.500 & 1.650 & 0.0287-0.0292\\
	\cline{3-5}
	Neutron star & 6.0  & 0.500 & 1.647-1.654 & 0.0290\\
	\cline{3-5}
	&   & 0.503-0.496 & 1.65 & 0.0290\\
	\hline
	 &  & 0.600 & 59.0 & (1.93-2.06)$\times10^{-5}$\\
	\cline{3-5}
	White dwarf & 5900.0 & 0.600 & 58.7-59.3 & 2.00$\times10^{-5}$\\
	\cline{3-5}
	&   & 0.585-0.610 & 59.0 & 2.00$\times10^{-5}$\\
	\hline
	\end{tabular}
\end{table}

\subsection{Around a WD}
\label{sect:wd_hydrodynamics}
Here we describe the hydrodynamics of the accretion flow around a WD which is assumed to be an MS with typical observed values of mass 0.8$M_\odot$ (\cite{Zorotovic2020}) with radius 0.01$R_\odot$ (\cite{Chandrasekhar1935,  Parsons2017}). Conversion of this radius in unit of $GM/c^2$ gives $a \sim$ 5900. Here we have assumed the eccentricity of the WD is 0.6. \cite{Balman2012, Balman2020} gives the transition radii (from standard SSD to advective disk) for quiescent dwarf novae (as in low state systems) that lie between (3-10)$\times 10^9$ cm (1-6 mHz break frequencies), which in units of $R_g$ becomes around (25000-85000) $R_g$. This indicates the possibility of presence of advective flow for a sufficiently large radial range which is required to form outer as well as inner critical point simultaneously. As outer and inner critical points occur simultaneously, it opens up a new possibility of formation of shock within the flow for white dwarf accretion as presence of two critical points together is a necessary condition for shock formation!!. Though it is not conventional, we used dimensionless unit to present length scales of white dwarf.
 
\begin{figure}

\includegraphics[scale=0.8]{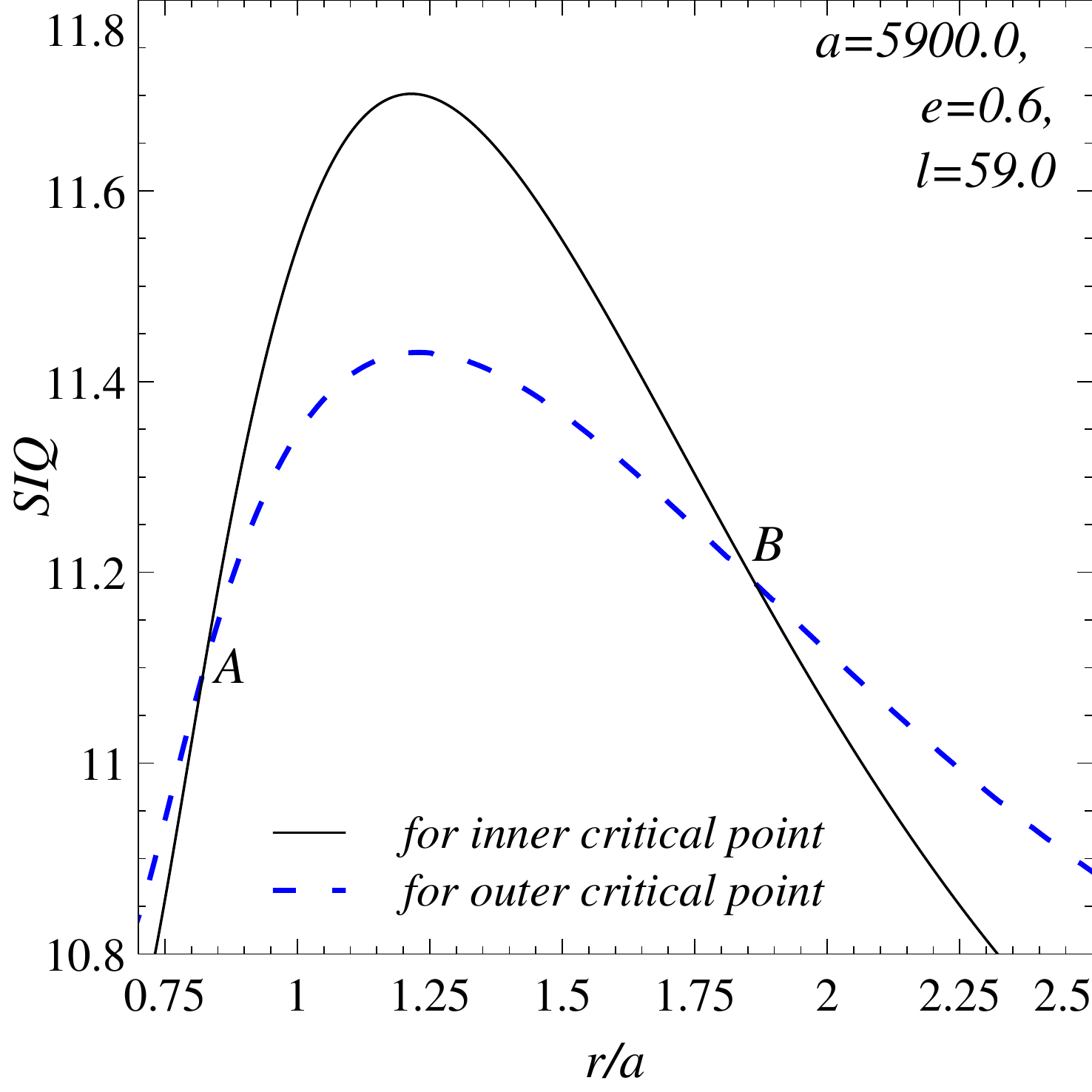}
\caption{Variation of shock invariant quantity with radial distance for accretion flow with $E_\mathrm{c}$=$2.0\times10^{-5}$ and $l$=59.0 around white dwarf. Two curves are for accretion through outer and inner critical point. The intersection points (A and B) of the two curves indicate the possible shock locations where inner one (A) is unstable and outer one (B) is representing the actual location of shock}
\label{SIQ_WD}
\end{figure}

\begin{figure*}

\includegraphics{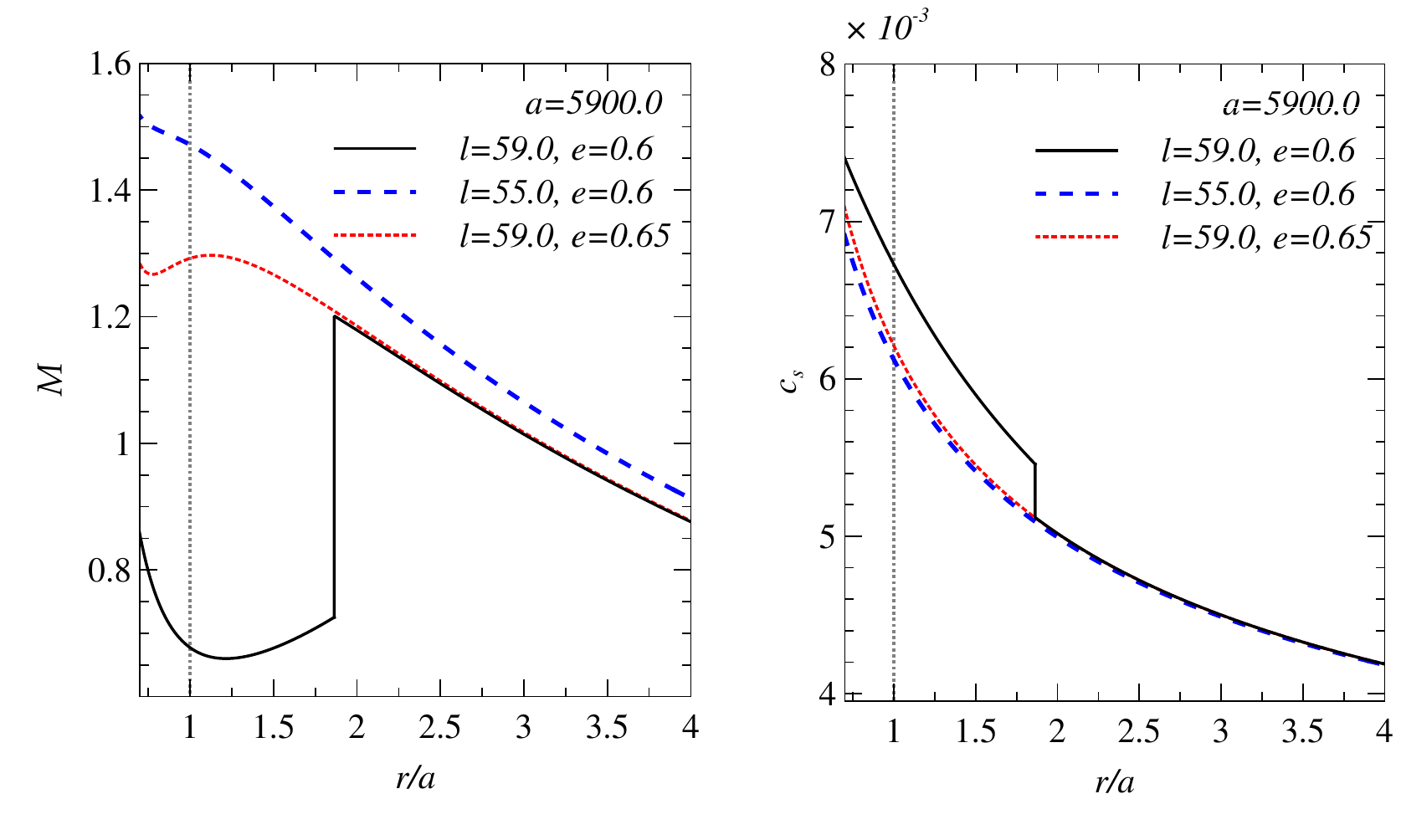}
\caption{Variation of Mach number ($\mathcal{M}$) and sound speed ($c_\mathrm{s}$) with radius same as Fig. {$\ref{NS hydro with without shock}$} but here compact object is white dwarf. Here combination of $l$=59.0 and $e$=0.6 gives $r_\mathrm{c,in}/a$=0.65 and $r_\mathrm{c,out}/a$=3.6 ($\sim$ 0.00017 Astronomical units) for $E_\mathrm{c}$=$2.0\times10^{-5}$ which makes shock possible at $r_\mathrm{sh}/a$=1.86. The plot is extended till inner critical point for shock solution. The vertical dotted line $r/a$=1 indicates the surface of the white dwarf.}

\label{WD hydro}
\end{figure*}

Here we have followed the same procedure as like for NS. Due to the larger size of WD, $l$ also increases largely as Keplerian angular momentum is proportional to $\sqrt r$. Also from equation (\ref{energy at critical point}) it is quite clear that the energy at critical point will reduce drastically due to the reduction in gravitational potential at larger distance. For advective flow with $E_\mathrm{c}$= $2.0\times10^{-5}$, $l$=59.0 around WD with $e$=0.6, there is simultaneous occurrence of outer as well as inner critical point. In Fig. \ref{SIQ_WD}, we have overplotted the $SIQ$ for two cases when accretion happens through outer as well as inner critical point. The outer intersection point (B) gives our desired shock location. We have shown radial profiles for $\mathcal{M}$ and $c_\mathrm{s}$ for $E_\mathrm{c}$= $2.0\times10^{-5}$ for three different sets: $l$=59.0, $e$=0.6; $l$=55.0, $e$=0.6; $l$=59.0, $e$=0.65 in Fig. \ref{WD hydro}. Only for ($l$=59.0, $e$=0.6) shock occurs, and there is a steep increase in $c_\mathrm{s}$ and a decrease in $\mathcal{M}$. For other two sets, shock disappears and there is gradual increase of $\mathcal{M}$ and $c_\mathrm{s}$ as matter goes inward through accretion. As the post shock region is hotter than unshocked flow, the post shock region will be puffed up and inflated as shown in Fig. 1 of \citealt{Chakrabarti1995}. For the shock solution, the outer and inner critical points occur at $r/a$=3.6 and 0.65 which makes the shock possible at $r/a$=1.86. Although, the inner critical point lies within the WD, the flow will remain undisturbed as the shock occurs within the flow outside of WD surface. This indicates again that shock can occur for only certain range of parameters. For a specific value of $E_\mathrm{c}$, only certain range of $l$ and $ae$ values will result in shock.  One set of values of different parameters which makes the shock possible for accretion onto a WD with $a$=5900.0 is tabulated in Table (\ref{range for shock}). However, it is evident as well as worth to mention that if we can vary all the parameters instaed of fixing two as tabulated in Table (\ref{range for shock}), the range of different parameters becomes quite large for which shock is possible. For accretion onto an WD with $a$=5900.0, shock will be possible by tuning the values of different parameters from range of $e$=0.2-1.0, $l_c$=32.0-76.0 and $E_c$=(1-8)$\times 10^{-5}$. Values of any parameter lying outside of this range will not be able to make the shock happen. Fig. \ref{WD hydro} is extended till the inner critical point for shock solution. The vertical dotted line indicates the location of the surface where the flow will stop. The angular momentum value increases in comparison with NS as the value of $a$ increases. Larger angular momentum is required to make effect at larger distance from central accretor.
 
Here, the result is more robust as hardly any general relativistic effect will come in the picture and also formation of shock is not possible within accretion flow in Newtonian potential of a spherical body. So, new phenomena arise only due to deformation of the shape of the accreting white dwarf. To the best of our knowledge we suppose, for the first time we are reporting this kind of shock formation for white dwarf accretion. As the formation of sub-Keplerian advective flow up to the surface of the white dwarf is possible for non-magnetic CVs, only spinning of WD should be sufficient to deform it. This is justified in discussion section further from observational point of view.

\section{Discussion}
We have discussed the hydrodynamics around an NS, as well as around a WD.
However, we must emphasize that the Newtonian description of the hydrodynamics is more suitable to describe the accretion physics around a WD compared to that around an NS; where general relativistic effects are important. 
Our proposed model is specifically applicable to the non-magnetic Cataclysmic Variables (CVs) where stellar magnetic field has a negligible effect on the accretion, unlike polars or intermediate polars (for a review see \citealt{Mukai2017}).
Observations reveal, more than $70$\% of the CVs are non-magnetic (\citealt{Ferrario2015, Ferrario2020}).

The non-magnetic CVs are broadly classified into the nova-likes (NLs) and dwarf novae (DNe). NLs spend most of their time in a high state (i.e high mass accretion rate) with UV emission predominantly originating from the disc (\citealt{Ladous1991}); while DNe are observed mostly in the quiescent  state (low mass accretion rate) (\citealt{Hack1993}). However, DNe show periodic outbursts (when mass accretion rate rises) with disk dominated optical and UV emission.

\subsection{Accretion with suitable parameters will give rise to shocks}
\label{sect:restricted_parameter_space}
Fig. (\ref{lambda sets}) and Fig. (\ref{ae sets}) show that there is a specific range of $l_c$ and $ae$ values for which simultaneous occurrence of two X-type critical points is possible for the accretion flow onto an NS and a WD. In the presence of two X-type critical points, shock becomes possible. Once $l_c$ and $ae$ are fixed, from the `swallow-tail' diagram (Fig. (\ref{Ec meuc with eccentricity})) we can estimate the energy range for which shock is possible. And finally matching the shock invariant quantity (equation (\ref{shock invariant quantity})), we find the exact energy of the flow which is required to make the shock possible for a suitable $l_c$ and $ae$. One set of values of different parameters suitable for shock for an NS having $a$=6.0 and for a WD having $a$=5900.0 is tabulated in Table (\ref{range for shock}). Once values of $a$, $e$, $l_c$ are fixed, there is a range of $E_c$ for which shock will occur. This is true for other parameters also. Always there is a range in value of one parameter when other parameters are fixed. It is quite encouraging that shock is possible for a wide range of eccentricity ($e$=0.2-1.0) of NS and WD. Therefore, in summary, accretion flow having suitable specific angular momentum and energy (here $l_c$ and $E_c$ mimics these values due to inviscid nature of the flow) will make shock when accreted to a suitable accretor (suitable $ae$ value). If conditions are suitable (appropriate values of parameters for the shock), a shocked solution is more likely to happen in the course of the accretion process instead of a shock free solution due to the nature of entropy which preferably increases.

\subsection{Is sufficient deformation of WD possible?}

Before drawing any connection of the described model with the observed phenomena in non-magnetic CVs, we would like to check the justification of the considered stellar deformation quantified by the parameter $e$. We find that shock is possible for a wide range of eccentricity ($e$=0.2-1.0) of NS and WD if values of other parameters are chosen suitably. As the $e$ of the accretor decreases, the required $l_c$ for the shock also decreases i.e. for a accretor which is more spherical (lower $e$ value, $e$=0 for a sphere), more sub-Keplerian flow (lesser angular momentum) is required to make the shock possible. Within the limited observed samples it has been estimated that the spin frequency $\Omega_*$ of many non-magnetic CVs lies in the range ($0.01 - 0.3$) $\Omega_K$ (\citealt{1999PASP..111..532S, Godon_Sion2012}), where $\Omega_K$ is the Keplerian angular velocity at the surface of the WD. Such values of $\Omega_*$ are good enough to generate the eccentricity $e$ ($\sim 0.2-0.4$) (equation (7.3.18) of \citealt{Shapiro1983}, \citealt{Chandrasekhar1969}) and even more oblate with low density of the spheroid. The possibility of shock for any $e\geq$ 0.2 (with appropriate $l_c$) indicates the applicability of our model for the WDs with observed rotational frequencies.

\subsection{Advective flows in CVs}
\label{sect:adaf_cv}

The UV spectra from the disc-dominated non-magnetic CVs (high state CVs) are generally modeled using the standard disc model (\citealt{Wade1998}). 
However, a significant fraction of the theoretical spectra were found to be too blue compared to the observed UV spectra (\citealt{Linnell2005, Puebla2007, Godon2017}).

Because of the presence of the hard surface in the WD, a boundary layer (BL) is expected to form between the stellar surface and the Keplerian disc (\citealt{Pringle1981, Frank2002}).
BL is predicted to be optically thick and to emit soft X-rays during high accretion states (high states in NLs and outbursts in DNe) (\citealt{Narayan1993, Popham1995}), while hard X-ray is expected from an optically thin BL region during quiescence (\citealt{Pringle1979, Narayan1993}). Although hard X-ray is observed in quiescent sources as expected (\citealt{Szkody2002, Pandel2005}), sources also exhibit optically thin hard X-ray emission in the high accretion states (\citealt{Teeseling1996, Balman2014, Balman2020}).

Inadequacy of the standard disc model in explaining observed UV spectra in high states  (\citealt{Puebla2007, Godon2017}) and presence of the optically thin hard X-rays in low (quiescent DNe) as well as in high accretion states as in NLs and outbursting DNe, is best explained in the context of radiatively inefficient hot advective accretion flows (ADAF-like) that exist in non-magnetic CVs (\citealt{Balman2012, Balman2020} and references therein). It is detected that an outer optically thick Keplerian disc is truncated to optically thin hot advective flow in the quiescent and outbursting DNe (\citealt{Balman2020} cf. Sec. 3). The outer Keplerian disc moves inward as mass accretion rate increases during the outbursts (\citealt{Balman2012, Balman2019}). Therefore a natural expectation is that a hot advective flow is supposed to exist both in low and in high accretion states of non-magnetic CVs. 

\subsection{Plausible connection of shock in advective flows to observables}
\label{section for connecting shock with observables}
The spectra from accreting WDs used to be modeled as multi-temperature isobaric cooling flow (power law and/or Bremsstrahlung) (mkcflow/CEVMKL in XSPEC or using two/more MEKAL components) (\citealt{Mukai2017} and references therein). In this model, the emission measure at each temperature is proportional to the time the cooling gas remains at that temperature. For most of the accreting WDs, multi-MEKAL component significantly improves the spectra in comparison with single component (\citealt{Mauche2002, Pandel2005, Balman2014}). Moreover, Presence of non-equilibrium ionization conditions of the plasma and the emission lines (N, O, Ne, Mg, Si, Fe) are detected for several nonmagnetic CVs in the X-rays (\citealt{Schlegel2014, Balman2020}), however this is not explored in the present form of our model.

One of the characteristic property of advective flows is that the inward flow velocity increases very rapidly as it approaches the central accretor (\citealt{Chakrabarti1989, Narayan1994, Chakrabarti1996}). The rapid increase in radial velocity also seen in the shock-free solutions denoted by the short and long dashed curves in Figs. \ref{NS hydro with without shock} and \ref{WD hydro}. Large radial velocity close to the accretor gives very little time to the flow for emission. Occurrence of shock in the advective flow (as discussed in section \ref{sect:hydrodynamics}) leads to  the rise in temperature ($\sim c^2_s$) as well as an increase in density in the post-shock region. In addition, in the post-shock flow, the radial velocity decreases, leading to a longer time span of the flow at higher temperatures. The optically thin radiatively inefficient advective flow in nonmagnetic CVs will give rise to radiation in the hard X-rays as they approach the central WD. The occurrence of the shock will alter the intensity as well as the hardness of the spectra which can also explain the variation of the efficiency of X-ray luminosity at different accretion rates (see \cite{Balman2020} sec. 2.3, 3 for a discussion). Hard X-rays are mainly observed in quiescent as well as in high accretion rate CVs and most related accreting WD systems (\citealt{Teeseling1996, Szkody2002, Pandel2005, Balman2014, Balman2020}). On a note of caution, as the location of the truncation radius moves closer to the central accretor in high accretion states of non-magnetic CVs, leaving a small radial range for the advective flow to make the shock happen, the occurrence of shock will become less probable.

\subsection{caveats}

In this work, we have analysed the sonic point formation, shocks in a steady inviscid sub-Keplerian advective flow and  and the related astrophysical implications. Due to the inviscid nature of the flow, the energy and the specific angular momentum of the flow at critical point ($E_c$ and $l_c$) becomes the energy and angular momentum of the flow at all radial distances. Energy and specific angular momentum remains constant throughout the flow. It can not be true in reality.

The specific angular momentum value we are using for formation of shock in WD accretion is $l_c$ = 59. At the WD surface, the ratio of specific angular momentum and Keplerian angular momentum becomes $\sim$ 0.77. Though it might be very low keeping the Keplerian disk in mind, it is customary to assume such low angular momentum value to do the analyses we have done in the sub-Keplerian advective flow (\citealt{Chakrabarti1989, Das2001, Mukhopadhyay2003, Palit2019, Dihingia2020}).

To solve the dynamics, we have used polytropic equation of state of the flow instead of solving energy equation. Also we have not incorporated any radiative cooling in the model. Therefore, it is not possible to comment anything on the actual spectra or luminosity generated due to the flow. In near future, we would like to do time-dependent calculations of the magnetized advective flow including radiative cooling and by solving the energy equation explicitly.

\section{Summary}
In this paper, we have studied the hydrodynamics of an optically thin advective flow around a deformed (from spherical shape) compact object which has a finite surface e.g., an NS or a WD. We treat the deformed star as a Maclaurin Spheroid (MS). Treating compact object as an MS opens up a new possibility for formation of Rankine-Hugoniot shock which is not possible for accretion around a spherical accretor without considering GR effects or effects of the hard surface of the accretor. To the best of our knowledge, for the first time we are reporting the possibility of occurrence of a shock in the radiatively inefficient hot advective flow around a WD. As far as we are concerned to understand the flow hydrodynamics around the non-magnetic CVs, our findings are robust as flow around a WD hardly deviates from the Newtonian regime. Although we have assumed inviscid flow and have not included radiative cooling for this preliminary study, we believe our findings will remain same qualitatively. We propose that some observational features e.g., high significance of fitting the spectra with multi-temperature plasma models rather than single temperature models, and variable efficiency of X-ray emission (X-ray luminosity in comparison with optical and UV luminosity of the disk) in nonmagnetic cataclysmic variables can be explained by the presence of shock in the sub-Keplerian advective flow.

\section*{Acknowledgements}

We would like to thank Banibrata Mukhopadhyay, Vikram Rana and Tushar Mondal for their useful discussion. We are very much thankful to the anonymous referee for insightful suggestions and referring to appropriate references. This work is partly supported by the fund of DST INSPIRE fellowship belonging to SRD.
\bibliography{macl_ref1}{}

\begin{thebibliography}{}
\expandafter\ifx\csname natexlab\endcsname\relax\def\natexlab#1{#1}\fi
\providecommand{\url}[1]{\href{#1}{#1}}
\providecommand{\dodoi}[1]{doi:~\href{http://doi.org/#1}{\nolinkurl{#1}}}
\providecommand{\doeprint}[1]{\href{http://ascl.net/#1}{\nolinkurl{http://ascl.net/#1}}}
\providecommand{\doarXiv}[1]{\href{https://arxiv.org/abs/#1}{\nolinkurl{https://arxiv.org/abs/#1}}}

\bibitem[{{Abramowicz} {et~al.}(1988){Abramowicz}, {Czerny}, {Lasota}, \&
  {Szuszkiewicz}}]{Abramowicz1988}
{Abramowicz}, M.~A., {Czerny}, B., {Lasota}, J.~P., \& {Szuszkiewicz}, E. 1988,
  \apj, 332, 646, \dodoi{10.1086/166683}

\bibitem[{{Amsterdamski} {et~al.}(2002){Amsterdamski}, {Bulik},
  {Gondek-Rosi{\'n}ska}, \& {Klu{\'z}niak}}]{Amsterdamski2002}
{Amsterdamski}, P., {Bulik}, T., {Gondek-Rosi{\'n}ska}, D., \& {Klu{\'z}niak},
  W. 2002, \aap, 381, L21, \dodoi{10.1051/0004-6361:20011555}

\bibitem[{{Balman}(2019)}]{Balman2019}
{Balman}, {\c{S}}. 2019, Astronomische Nachrichten, 340, 296,
  \dodoi{10.1002/asna.201913613}

\bibitem[{{Balman}(2020)}]{Balman2020}
---. 2020, Advances in Space Research, 66, 1097,
  \dodoi{10.1016/j.asr.2020.05.031}

\bibitem[{{Balman} {et~al.}(2014){Balman}, {Godon}, \& {Sion}}]{Balman2014}
{Balman}, {\c S}., {Godon}, P., \& {Sion}, E.~M. 2014, \apj, 794, 84,
  \dodoi{10.1088/0004-637X/794/1/84}

\bibitem[{{Balman} \& {Revnivtsev}(2012)}]{Balman2012}
{Balman}, {\c S}., \& {Revnivtsev}, M. 2012, \aap, 546, A112,
  \dodoi{10.1051/0004-6361/201219469}

\bibitem[{{Bhattacharjee} \& {Chakrabarti}(2019)}]{Bhattacharjee2019}
{Bhattacharjee}, A., \& {Chakrabarti}, S.~K. 2019, \apj, 873, 119,
  \dodoi{10.3847/1538-4357/ab074a}

\bibitem[{{Bondi}(1952)}]{Bondi1952}
{Bondi}, H. 1952, \mnras, 112, 195, \dodoi{10.1093/mnras/112.2.195}

\bibitem[{{Chakrabarti} \& {Titarchuk}(1995)}]{Chakrabarti1995}
{Chakrabarti}, S., \& {Titarchuk}, L.~G. 1995, \apj, 455, 623,
  \dodoi{10.1086/176610}

\bibitem[{{Chakrabarti}(1989)}]{Chakrabarti1989}
{Chakrabarti}, S.~K. 1989, \apj, 347, 365, \dodoi{10.1086/168125}

\bibitem[{{Chakrabarti}(1990)}]{Chakrabarti1990}
---. 1990, {Theory of Transonic Astrophysical Flows} (World Scientific
  Publishing Co), \dodoi{10.1142/1091}

\bibitem[{{Chakrabarti}(1996)}]{Chakrabarti1996}
---. 1996, \apj, 464, 664, \dodoi{10.1086/177354}

\bibitem[{Chandrasekhar(1935)}]{Chandrasekhar1935}
Chandrasekhar, S. 1935, Monthly Notices of the Royal Astronomical Society, 95,
  207, \dodoi{10.1093/mnras/95.3.207}

\bibitem[{{Chandrasekhar}(1969)}]{Chandrasekhar1969}
{Chandrasekhar}, S. 1969, {Ellipsoidal figures of equilibrium}

\bibitem[{Chatterjee {et~al.}(2018)Chatterjee, Chakrabarti, Ghosh, \&
  Garain}]{Chatterjee2018}
Chatterjee, A., Chakrabarti, S.~K., Ghosh, H., \& Garain, S.~K. 2018, Monthly
  Notices of the Royal Astronomical Society, 478, 3356,
  \dodoi{10.1093/mnras/sty1054}

\bibitem[{{Das} {et~al.}(2001){Das}, {Chattopadhyay}, \&
  {Chakrabarti}}]{Das2001}
{Das}, S., {Chattopadhyay}, I., \& {Chakrabarti}, S. i.~K. 2001, \apj, 557,
  983, \dodoi{10.1086/321692}

\bibitem[{{Das} \& {Mukhopadhyay}(2015)}]{Das2015}
{Das}, U., \& {Mukhopadhyay}, B. 2015, \jcap, 2015, 016,
  \dodoi{10.1088/1475-7516/2015/05/016}

\bibitem[{{Dhang} {et~al.}(2016){Dhang}, {Sharma}, \&
  {Mukhopadhyay}}]{Dhang2016}
{Dhang}, P., {Sharma}, P., \& {Mukhopadhyay}, B. 2016, \mnras, 461, 2426,
  \dodoi{10.1093/mnras/stw1480}

\bibitem[{{Dhang} {et~al.}(2018){Dhang}, {Sharma}, \&
  {Mukhopadhyay}}]{Dhang2018}
---. 2018, \mnras, 476, 3310, \dodoi{10.1093/mnras/sty488}

\bibitem[{{Dihingia} {et~al.}(2020){Dihingia}, {Das}, {Prabhakar}, \& {Mand
  al}}]{Dihingia2020}
{Dihingia}, I.~K., {Das}, S., {Prabhakar}, G., \& {Mand al}, S. 2020, \mnras,
  496, 3043, \dodoi{10.1093/mnras/staa1687}

\bibitem[{{Ferrario} {et~al.}(2015){Ferrario}, {de Martino}, \&
  {G{\"a}nsicke}}]{Ferrario2015}
{Ferrario}, L., {de Martino}, D., \& {G{\"a}nsicke}, B.~T. 2015, \ssr, 191,
  111, \dodoi{10.1007/s11214-015-0152-0}

\bibitem[{{Ferrario} {et~al.}(2020){Ferrario}, {Wickramasinghe}, \&
  {Kawka}}]{Ferrario2020}
{Ferrario}, L., {Wickramasinghe}, D., \& {Kawka}, A. 2020, Advances in Space
  Research, 66, 1025, \dodoi{10.1016/j.asr.2019.11.012}

\bibitem[{{Frank} {et~al.}(2002){Frank}, {King}, \& {Raine}}]{Frank2002}
{Frank}, J., {King}, A., \& {Raine}, D.~J. 2002, {Accretion Power in
  Astrophysics: Third Edition}, 398

\bibitem[{{Godon} {et~al.}(2017){Godon}, {Sion}, {Balman}, \&
  {Blair}}]{Godon2017}
{Godon}, P., {Sion}, E.~M., {Balman}, {\c S}., \& {Blair}, W.~P. 2017, \apj,
  846, 52, \dodoi{10.3847/1538-4357/aa7f71}

\bibitem[{{Godon} {et~al.}(2012){Godon}, {Sion}, {Levay}, {Linnell}, {Szkody},
  {Barrett}, {Hubeny}, \& {Blair}}]{Godon_Sion2012}
{Godon}, P., {Sion}, E.~M., {Levay}, K., {et~al.} 2012, \apjs, 203, 29,
  \dodoi{10.1088/0067-0049/203/2/29}

\bibitem[{Hack \& La~Dous(1993)}]{Hack1993}
Hack, M., \& La~Dous, C. 1993, {Cataclysmic variables and related objects},
  Vol. 507 (National Aeronautics and Space Administration, Scientific and
  Technical~…)

\bibitem[{{Haskell} {et~al.}(2008){Haskell}, {Samuelsson}, {Glampedakis}, \&
  {Andersson}}]{2008MNRAS.385..531H}
{Haskell}, B., {Samuelsson}, L., {Glampedakis}, K., \& {Andersson}, N. 2008,
  \mnras, 385, 531, \dodoi{10.1111/j.1365-2966.2008.12861.x}

\bibitem[{{Iwakami} {et~al.}(2009){Iwakami}, {Kotake}, {Ohnishi}, {Yamada}, \&
  {Sawada}}]{Iwakami2009}
{Iwakami}, W., {Kotake}, K., {Ohnishi}, N., {Yamada}, S., \& {Sawada}, K. 2009,
  \apj, 700, 232, \dodoi{10.1088/0004-637X/700/1/232}

\bibitem[{{Klu{\'z}niak} \& {Rosi{\'n}ska}(2013)}]{Kluzniak2013}
{Klu{\'z}niak}, W., \& {Rosi{\'n}ska}, D. 2013, \mnras, 434, 2825,
  \dodoi{10.1093/mnras/stt1185}

\bibitem[{{Komatsu} {et~al.}(1989){Komatsu}, {Eriguchi}, \&
  {Hachisu}}]{Komatsu1989}
{Komatsu}, H., {Eriguchi}, Y., \& {Hachisu}, I. 1989, \mnras, 237, 355,
  \dodoi{10.1093/mnras/237.2.355}

\bibitem[{{La Dous}(1991)}]{Ladous1991}
{La Dous}, C. 1991, \aap, 252, 100

\bibitem[{Landau(1987)}]{landau1987fluid}
Landau, L. 1987, Fluid Mechanics: Landau and Lifshitz Course of Theoretical
  Physics, vol. 6,  Pergamon Press, Oxford

\bibitem[{{Linnell} {et~al.}(2005){Linnell}, {Szkody}, {G{\"a}nsicke}, {Long},
  {Sion}, {Hoard}, \& {Hubeny}}]{Linnell2005}
{Linnell}, A.~P., {Szkody}, P., {G{\"a}nsicke}, B., {et~al.} 2005, \apj, 624,
  923, \dodoi{10.1086/429143}

\bibitem[{{Mauche} \& {Mukai}(2002)}]{Mauche2002}
{Mauche}, C.~W., \& {Mukai}, K. 2002, \apjl, 566, L33, \dodoi{10.1086/339454}

\bibitem[{{Medvedev} \& {Menou}(2002)}]{Medvedev2002}
{Medvedev}, M.~V., \& {Menou}, K. 2002, \apjl, 565, L39, \dodoi{10.1086/339053}

\bibitem[{{Mishra} \& {Vaidya}(2015)}]{Mishra2015}
{Mishra}, B., \& {Vaidya}, B. 2015, \mnras, 447, 1154,
  \dodoi{10.1093/mnras/stu2468}

\bibitem[{{Molteni} {et~al.}(1996){Molteni}, {Sponholz}, \&
  {Chakrabarti}}]{Molteni1996}
{Molteni}, D., {Sponholz}, H., \& {Chakrabarti}, S.~K. 1996, \apj, 457, 805,
  \dodoi{10.1086/176775}

\bibitem[{{Molteni} {et~al.}(1999){Molteni}, {T{\'o}th}, \&
  {Kuznetsov}}]{Molteni1999}
{Molteni}, D., {T{\'o}th}, G., \& {Kuznetsov}, O.~A. 1999, \apj, 516, 411,
  \dodoi{10.1086/307079}

\bibitem[{{Mukai}(2017)}]{Mukai2017}
{Mukai}, K. 2017, \pasp, 129, 062001, \dodoi{10.1088/1538-3873/aa6736}

\bibitem[{{Mukhopadhyay}(2002)}]{Mukhopadhyay2002}
{Mukhopadhyay}, B. 2002, \apj, 581, 427, \dodoi{10.1086/344227}

\bibitem[{{Mukhopadhyay}(2003)}]{Mukhopadhyay2003}
---. 2003, \apj, 586, 1268, \dodoi{10.1086/367830}

\bibitem[{{Nakayama}(1992)}]{Nakayama1992}
{Nakayama}, K. 1992, \mnras, 259, 259, \dodoi{10.1093/mnras/259.2.259}

\bibitem[{{Narayan} {et~al.}(1998){Narayan}, {Mahadevan}, \&
  {Quataert}}]{Narayan1998}
{Narayan}, R., {Mahadevan}, R., \& {Quataert}, E. 1998, in Theory of Black Hole
  Accretion Disks, ed. M.~A. {Abramowicz}, G.~{Bj{\"o}rnsson}, \& J.~E.
  {Pringle}, 148--182

\bibitem[{{Narayan} \& {Popham}(1993)}]{Narayan1993}
{Narayan}, R., \& {Popham}, R. 1993, \nat, 362, 820, \dodoi{10.1038/362820a0}

\bibitem[{{Narayan} \& {Yi}(1994)}]{Narayan1994}
{Narayan}, R., \& {Yi}, I. 1994, \apjl, 428, L13, \dodoi{10.1086/187381}

\bibitem[{{Narayan} \& {Yi}(1995)}]{Narayan1995}
---. 1995, \apj, 452, 710, \dodoi{10.1086/176343}

\bibitem[{{Nobuta} \& {Hanawa}(1994)}]{Nobuta1994}
{Nobuta}, K., \& {Hanawa}, T. 1994, \pasj, 46, 257

\bibitem[{{Novikov} \& {Thorne}(1973)}]{Novikov1973}
{Novikov}, I.~D., \& {Thorne}, K.~S. 1973, in Black Holes (Les Astres Occlus),
  ed. C.~{Dewitt} \& B.~S. {Dewitt}, 343--450

\bibitem[{{Ostriker} \& {Bodenheimer}(1968)}]{1968ApJ...151.1089O}
{Ostriker}, J.~P., \& {Bodenheimer}, P. 1968, \apj, 151, 1089,
  \dodoi{10.1086/149507}

\bibitem[{{Ostriker} \& {Hartwick}(1968)}]{1968ApJ...153..797O}
{Ostriker}, J.~P., \& {Hartwick}, F.~D.~A. 1968, \apj, 153, 797,
  \dodoi{10.1086/149706}

\bibitem[{{Paczy{\'n}sky} \& {Wiita}(1980)}]{Paczynski1980}
{Paczy{\'n}sky}, B., \& {Wiita}, P.~J. 1980, \aap, 88, 23

\bibitem[{Palit {et~al.}(2019)Palit, Janiuk, \& Sukova}]{Palit2019}
Palit, I., Janiuk, A., \& Sukova, P. 2019, Monthly Notices of the Royal
  Astronomical Society, 487, 755, \dodoi{10.1093/mnras/stz1296}

\bibitem[{{Pandel} {et~al.}(2005){Pandel}, {C{\'o}rdova}, {Mason}, \&
  {Priedhorsky}}]{Pandel2005}
{Pandel}, D., {C{\'o}rdova}, F.~A., {Mason}, K.~O., \& {Priedhorsky}, W.~C.
  2005, \apj, 626, 396, \dodoi{10.1086/429983}

\bibitem[{{Parsons} {et~al.}(2017){Parsons}, {G{\"a}nsicke}, {Marsh}, {Ashley},
  {Bours}, {Breedt}, {Burleigh}, {Copperwheat}, {Dhillon}, {Green}, {Hardy},
  {Hermes}, {Irawati}, {Kerry}, {Littlefair}, {McAllister}, {Rattanasoon},
  {Rebassa-Mansergas}, {Sahman}, \& {Schreiber}}]{Parsons2017}
{Parsons}, S.~G., {G{\"a}nsicke}, B.~T., {Marsh}, T.~R., {et~al.} 2017, \mnras,
  470, 4473, \dodoi{10.1093/mnras/stx1522}

\bibitem[{{Popham} \& {Narayan}(1995)}]{Popham1995}
{Popham}, R., \& {Narayan}, R. 1995, \apj, 442, 337, \dodoi{10.1086/175444}

\bibitem[{{Pringle}(1981)}]{Pringle1981}
{Pringle}, J.~E. 1981, \araa, 19, 137,
  \dodoi{10.1146/annurev.aa.19.090181.001033}

\bibitem[{{Pringle} \& {Savonije}(1979)}]{Pringle1979}
{Pringle}, J.~E., \& {Savonije}, G.~J. 1979, \mnras, 187, 777,
  \dodoi{10.1093/mnras/187.4.777}

\bibitem[{{Puebla} {et~al.}(2007){Puebla}, {Diaz}, \& {Hubeny}}]{Puebla2007}
{Puebla}, R.~E., {Diaz}, M.~P., \& {Hubeny}, I. 2007, \aj, 134, 1923,
  \dodoi{10.1086/522112}

\bibitem[{{Rajesh} \& {Mukhopadhyay}(2010)}]{Rajesh2010}
{Rajesh}, S.~R., \& {Mukhopadhyay}, B. 2010, \mnras, 402, 961,
  \dodoi{10.1111/j.1365-2966.2009.15925.x}

\bibitem[{Schlegel {et~al.}(2014)Schlegel, Shipley, Rana, Barrett, \&
  Singh}]{Schlegel2014}
Schlegel, E.~M., Shipley, H.~V., Rana, V.~R., Barrett, P.~E., \& Singh, K.~P.
  2014, The Astrophysical Journal, 797, 38, \dodoi{10.1088/0004-637x/797/1/38}

\bibitem[{{Shakura} \& {Sunyaev}(1973)}]{Shakura1973}
{Shakura}, N.~I., \& {Sunyaev}, R.~A. 1973, \aap, 24, 337

\bibitem[{{Shapiro} \& {Teukolsky}(1983)}]{Shapiro1983}
{Shapiro}, S.~L., \& {Teukolsky}, S.~A. 1983, {Black holes, white dwarfs, and
  neutron stars: The physics of compact objects}

\bibitem[{{Sion}(1999)}]{1999PASP..111..532S}
{Sion}, E.~M. 1999, \pasp, 111, 532, \dodoi{10.1086/316361}

\bibitem[{{Subramanian} \& {Mukhopadhyay}(2015)}]{Subramanian2015}
{Subramanian}, S., \& {Mukhopadhyay}, B. 2015, \mnras, 454, 752,
  \dodoi{10.1093/mnras/stv1983}

\bibitem[{{Szkody} {et~al.}(2002){Szkody}, {Nishikida}, {Raymond}, {Seth},
  {Hoard}, {Long}, \& {Sion}}]{Szkody2002}
{Szkody}, P., {Nishikida}, K., {Raymond}, J.~C., {et~al.} 2002, \apj, 574, 942,
  \dodoi{10.1086/341006}

\bibitem[{{van Teeseling} {et~al.}(1996){van Teeseling}, {Beuermann}, \&
  {Verbunt}}]{Teeseling1996}
{van Teeseling}, A., {Beuermann}, K., \& {Verbunt}, F. 1996, \aap, 315, 467

\bibitem[{{Wade} \& {Hubeny}(1998)}]{Wade1998}
{Wade}, R.~A., \& {Hubeny}, I. 1998, \apj, 509, 350, \dodoi{10.1086/306496}

\bibitem[{{Yuan} {et~al.}(2005){Yuan}, {Cui}, \& {Narayan}}]{Yuan2005}
{Yuan}, F., {Cui}, W., \& {Narayan}, R. 2005, \apj, 620, 905,
  \dodoi{10.1086/427206}

\bibitem[{{Zorotovic} \& {Schreiber}(2020)}]{Zorotovic2020}
{Zorotovic}, M., \& {Schreiber}, M.~R. 2020, Advances in Space Research, 66,
  1080, \dodoi{10.1016/j.asr.2019.08.044}

\end{thebibliography}
\bibliographystyle{aasjournal}
\end{document}